# Size-adjustable Ring-shape Photoacoustic Tomography System

Daohuai Jiang, Yifei Xu, Hengrong Lan, Feng Gao, and Fei Gao, *Member, IEEE*

*Abstract*—Photoacoustic tomography (PAT) combines the advantages of the spectroscopic optical absorption contrast and acoustic resolution with deep penetration, and becomes an important novel biomedical imaging technology for scientific research and clinical diagnosis. In this paper, an imaging size-adjustable PAT system is proposed for clinical applications, which can adapt for different size imaging targets. Comparing with the conventional PAT system setup which with a fixed radius ring shape ultrasound transducer (UT) array, the proposed system, which is based on sectorial ultrasound transducer array (SUTA), is more flexible for different size targets' imaging. There are 32 elements for each SUTA, and four SUTAs form a 128-channel UT array for photoacoustic wave detection. The four SUTAs are controlled by four stepper motors, and change the SUTAs distribute positon that adapt for imaging applications. The radius of the proposed system imaging region of interest (ROI) can be adjusted from 50 mm to 90 mm, which is much more flexible than the conventional full ring UT array PAT system. The simulation results generated by the MATLAB k-wave toolbox very well demonstrate the feasibility of the proposed system. To further validate the proposed system for size-adjustable imaging, a vascular mimicking phantom and ex-vivo pork breast with indocyanine green (ICG) injected are imaged to prove its feasibility for clinical applications.

## I. Introduction

In the female population, breast cancer is the most common malignancies with high mortality rate [1]. Breast cancer, along with lung and bronchial cancers, are the two cancers with the highest number of new cases and deaths among women. Breast cancer accounts for 29% of all new cancer cases and 15% of all deaths in women, ranking first and second among all the cancer types, respectively. Between the 1980s and 1990s, the incidence of in situ breast cancer increased rapidly, largely due to the popularity of mammography screening, demonstrating that the use of advanced imaging techniques is highly beneficial for the early screening of breast cancer [2].

Over the past few decades, many medical imaging techniques have been used to detect breast cancer, with the more common ones being X-ray mammography, ultrasound imaging, and MRI. However, the clinical diagnosis of breast cancer still faces challenges due to the inherent shortcomings of these methods. X-ray mammography is the most commonly used screening methods, but its application is still limited due to ionizing radiation, lower sensitivity and specificity [3]. Ultrasound imaging is limited by its high false positive, lower sensitivity and specificity, and its reproducibility and reliability are questionable due to its high operator dependence [4, 5]. In addition, MRI has been shown to be more effective than X-ray mammography and ultrasonography, but suffers from its high cost and contrast agent requirements, which prevents its application from cancer screening widely [6, 7].

Clinical research has shown that early detection of breast cancer can significantly improve the 5-year survival rate of patients [3]. Considering the limitations of existing imaging modalities, a new imaging method that can be used for accurate early breast cancer screening is needed. Such new imaging methods need to overcome some of the shortcomings of current methods, including but not limited to: (1) No ionizing radiation; (2) high sensitivity and specificity; (3) good repeatability; (4) lower costs; (5) sufficient resolution; (6) fast imaging speed.

Photoacoustic tomography (PAT) is an emerging non-invasive biomedical imaging method [8-10]. The PAT system combines the advantages of both optical contrast and acoustic penetration to achieve high-resolution functional and molecular imaging in biological tissues with unprecedented centimeter-level depth [11, 12]. In recent years, more and more scientific studies have shown the potential of this technology to be used in pre-clinical and clinical applications [10, 13, 14]. PAT imaging is based on the photoacoustic effect. When a laser beam is shot on biological tissues, the optical energy is absorbed by the tissues and converted into heat energy. Then, the photoacoustic pressure caused by the thermal expansion of the biological tissue is generated and propagated in the tissue in the form of ultrasound. The ultrasonic transducer is coupled with the biological tissue through the coupling agent to receive

This research was funded by Natural Science Foundation of Shanghai (18ZR1425000), and National Natural Science Foundation of China (61805139). (Corresponding author: Fei Gao.)
Daohuai Jiang and Yifei Xu are equal to this work.
Daohuai Jiang (e-mail: jiangdh1@shanghaitech.edu.cn), Yifei Xu (e-mail: xuyf19971206@gmail.com) and Hengrong Lan (e-mail: lanhr@shanghaitech.edu.cn) are with the Hybrid Imaging System Laboratory (HISLab), Shanghai Engineering Research Center of Intelligent Vision and Imaging, School of Information Science and Technology, ShanghaiTech University, Shanghai 201210, China, with Chinese Academy of Sciences, Shanghai Institute of Microsystem and Information Technology, Shanghai 200050, China, and also with University of Chinese Academy of Sciences, Beijing 100049, China.
Feng Gao and Fei Gao are with the Hybrid Imaging System Laboratory (HISLab), Shanghai Engineering Research Center of Intelligent Vision and Imaging, School of Information Science and Technology, ShanghaiTech University, Shanghai 201210, China (e-mail: gaofeng@shanghaitech.edu.cn, gaofei@shanghaitech.edu.cn).

the PA signals, which can be used to reconstruct the optical absorption distribution inside the tissue.

The PAT method has been widely studied for the detection of breast cancer. The main difference between the existing photoacoustic imaging systems is the arrangement of the ultrasound transducers. The conventional design of breast photoacoustic imaging systems can be divided into three types: linear, planar, and hemispherical. Linear distribution systems typically use conventional handheld linear-array ultrasound probe with fixed optical fibers on either side of the transducer array for light illumination. This allows the breast to be illuminated and detected in the same direction [12]. This design allows for easy exploration of the area of interest, but the effectiveness is highly dependent on the operator's skill with limited ROI. The planar system clamps the breast with two flat plates, allowing detection with a two-dimensional array of transducers [15]. This method yields a larger field of view with fewer artifacts, but it is difficult to obtain a complete view near the chest wall. The hemispherical system allows for a complete 3D scan of the breast by mounting a dense array of ultrasound transducers in a bowl shape [16]. Designs that achieve similar results by rotating a linear array are also available [17]. This design provides the greatest number of detection angles and thus suffers the least from artifacts, and allows for a larger field of view. However, the use of a considerable number of transducers makes both data acquisition and processing challenges, often requiring longer reconstruction times and higher cost. Currently, clinical results have been achieved with photoacoustic detection of breast cancer, but several of the system designs mentioned above have some problems when considering specific applications for rapid screening of breast cancer. One disadvantage is that these systems do not take into account the diversity of breast sizes, which may severely hamper the performance of the breast imaging.

In this paper, we designed a size-adjustable SUTA-based rapid breast screening system, which has substantial improvement compared with our previous work [18]. The system uses four separated SUTAs that are position-adjusted by four stepper motors to provide high-resolution rapid screening, while adapting to a larger population. In this paper, we demonstrated the ROI size-adjustable PAT system both in simulation and phantom study. The proposed system can adapt for different-size imaging targets, and improve the flexibility of the PAT system for clinical applications. In this paper, the following section II will introduce the method of the SUTA-based ROI size-adjustable PAT system including the transducer array design, SUTA distributions and imaging reconstructions. In section III, the simulation and phantom study are executed to verify the feasibility of the size-adjustable PAT system. Section IV presents discussions and conclusions for the proposed PAT system.

II. METHOD

*A. Overview*

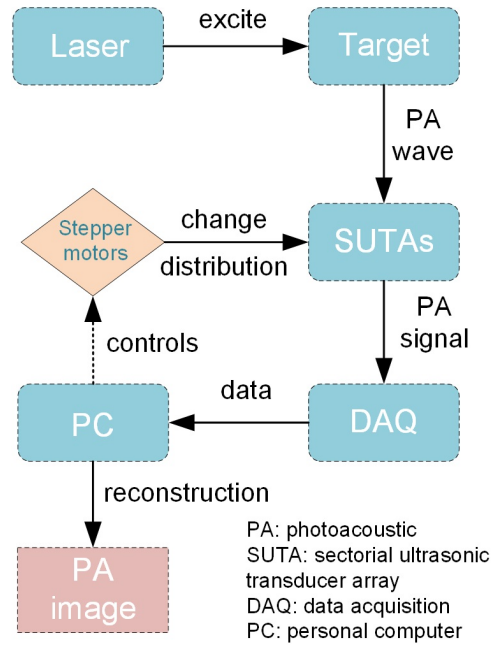

Fig. 1. The overview of the size-adjustable PAT system flow.

Fig. 1 shows the overview of the system flow, the laser source excites the imaging target and then generates PA wave; the computer controls the stepper motors that change the distributions of the four SUTAs; the PA signals received by the SUTAs are sampled by the data acquisition (DAQ); after that, the computer reconstructs the PA image by applying the back-projection algorithms.

The stepper motors drive the four SUTAs to move toward the imaging target, which ensures that the transducer can better receive the PA signals with tight enclosure. For image reconstruction, the coordinates of the ultrasonic transducer elements of the SUTAs are necessary. According to the stepper motors movements and the shape of SUTA, the coordinate of each transducer element can be uniquely calculated for image reconstruction. Therefore, the proposed system based on the SUTAs can realize size adjustable ROI for PAT imaging.

*B. Sectorial Ultrasound Transducer Array Design*

The ultrasound transducer is an essential part of the PAT system for PA wave detection [19]. To reconstruct a high-fidelity PA image, the PA signals should be received at full angles. Therefore, for the conventional PAT system setup, a ring-shaped UT array with full-angle UT elements distribution is preferred to receive the PA signals for image reconstruction. In this paper, we proposed a sectorial shape UT array for PA wave detection, and Fig. 2 shows 3D plot of the custom-designed SUTA structure. The PA signal is detected by four SUTAs with a 90-degree interval in a plane that forms a ring-shaped UT array. The four SUTAs distribution can be flexibly changed according to the imaging targets. The dimension of the SUTA is shown in Fig. 2. The SUTA with an axial symmetry shape has the length equal to the width of 90 mm, and the height is 20 mm. For each SUTA, there are 32 elements for ultrasound detection, which evenly distribute on the sector arc. Each element is with 2.317 mm width and 10 mm height. The degree of the sector arc is 70.8 for ultrasound detection. Therefore, by applying four SUTAs, it can form a 128-channel size-adjustable UT array.

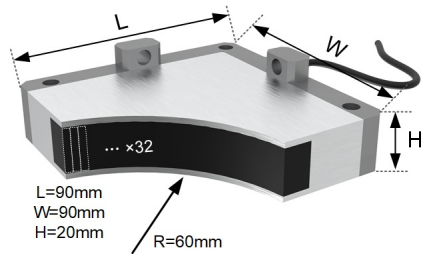

Fig. 2. The proposed SUTA schematic diagram. L: length; W: width; H: height; R: radius.

The photoacoustic signal is broadband signal, to get a more detailed reconstructed image, more frequency component capture is necessary. Therefore, four kinds of SUTA with different center frequency are designed in this work. We selected 7.5 MHz, 5 MHz, 2.5 MHz and 1 MHz as the ultrasound transducer center frequency with 60% bandwidth, which can cover the PA signals' spectrum for most PAT system applications.

*C. SUTA's Distribution with ROI size-adjustable*

In this work, the ultrasound detection part of the PAT system is the four SUTAs based 128-channel transducer array, each SUTA's position can be flexibly changed for better PA signal detection. Therefore, changing the four SUTAs distribute position for PA signal receiving can realize a ROI size-adjustable PAT imaging. As shown in Fig. 3(a)-(c), they demonstrate the three situations that the four SUTAs forms a 128-channel array for PA signals detection.

For the small target, the ultrasound transducer close to the PA wave source can get better detection sensitivity that avoids the severe acoustic attenuation of the propagation. Therefore, the four SUTAs close to the imaging center forms a small region for PA signals detection as shown in Fig. 3(a). Changing the SUTAs position can adept for different size imaging target, as shown in Fig. 3(b) and Fig. 3(c), which can be used for middle and large targets imaging. The proposed SUTA based ROI size-adjustable PAT system can realize the radius range from 50 mm to 90 mm PA signals detection with high flexibility for different size target imaging. However, the gap between SUTAs increases, along with the ROI size enlargement. Therefore, for large ROI imaging, there is still limited view issue, which can be solved by rotational compensation.

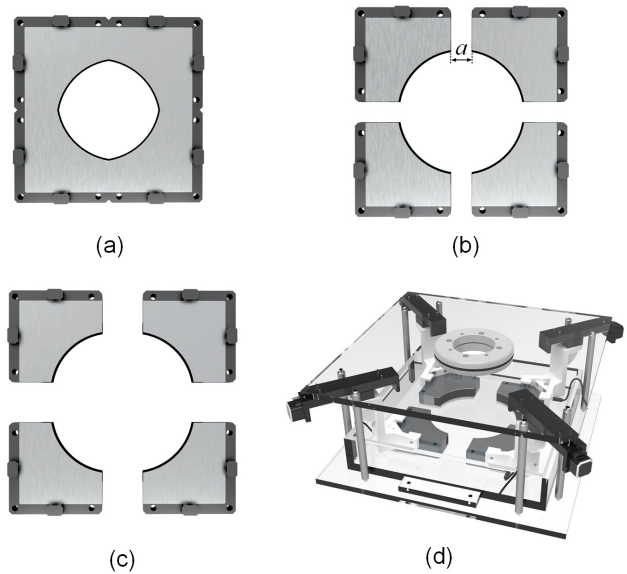

Fig. 3. The ROI size-adjustable PAT system with four SUTAs. (a) is the situation for small ROI with 50 mm radius, (b) forms a middle size region with 60 mm radius and (c) is for large ROI; (d) Illustration of the four SUTAs' distribution controlled by stepper motors.

*D. SUTA signal calibration and image reconstruction*

The four SUTAs are fixed to the stepper motors, and the movement of the motors is controlled to adjust the size of the imaging ROI. However, due to the inevitable errors in the distribution of the four SUTAs, both the hardware and the PA signals need to be corrected for better imaging quality. The distribute position error of the SUTAs can be divided into four cases, as shown in Fig. 4. Fig. 4(a) can be considered as the ideal distribution of the SUTA that the probe towards the imaging center without offset errors. Fig. 4(b) shows that the SUTA is oriented towards the imaging center, but with a backward offset error. Fig. 4(c) illustrates the case where the SUTA is not oriented towards the imaging center. And Fig. 4(d) shows the SUTA towards the imaging center with a forward offset error. All of the three kinds of errors in Fig. 4(b)-(d) should be calibrated before image reconstruction. All three types of errors mentioned above can be considered as signals with different delays compared to the signals received by the ideal SUTA distribution for the system. The errors illustrated in Fig. 4(b) can be corrected by moving the SUTA position closer to the imaging area or by shifting the signal forward. In Fig. 4(c), the SUTA is offset with rotation, and the transducer needs to be corrected to be oriented towards the imaging center. Alternatively, the received signal needs to be shifted according to its position to compensate for the error. For the error correction of Fig. 4(d), the SUTA has moved away from the imaging center, which can be calibrated by backward shifting the signals. To test and calibrate the errors, a black spot can be placed in the center of the imaging area. After laser excitation, the PA signals can be received by the SUTAs. According to the PA signals' delay analyses, the corresponding correction method can be obtained[20].

The data shifting bits and SUTA position error distance relationship can be calculated as:

$$n = \frac{d}{v} \cdot f_s \quad (1)$$

where $n$ stands for the bit number in the digtal domain of the shifting operation, $d$ is the error of the SUTAs spatial distribution between the calculate position coordinate, $v$ is the photoacoustic wave propagation speed in the imaging medium, the data acquisition sampling rate is $f_s$.

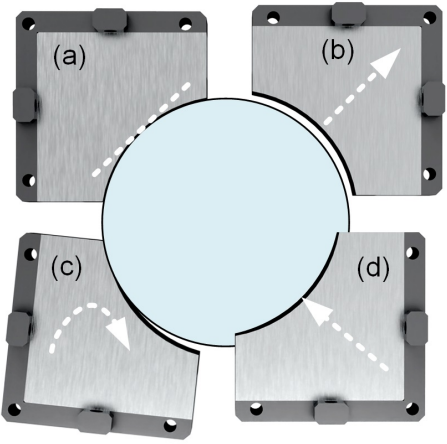

Fig. 4. The four SUTAs distribution diagram with position error. (a) the SUTA is distributed towards the center of the imaging area with no offset errors; (b) the SUTA is distributed towards the center of the imaging area with backward offset error; (c) the situation of SUTA is not oriented towards the center of the imaging area; (d) the SUTA is distributed towards the center of the imaging area with forwarding offset error.

After the PA signals corrections, the PA image can be reconstructed by the universal back-projection algorithms. The acoustic wave, p(r,t) at position r and time t, can be described as[11, 21]:

$$\left(\nabla^2 - \frac{1}{c^2}\frac{\partial^2}{\partial t^2}\right)p(r,t) = 0 \quad (2)$$

Subject to the initial conditions:

$$p(r,t)|_{t=0} = \frac{\beta c^2}{C_p}A(r); \quad \frac{\partial p(r,t)}{\partial t}|_{t=0} = 0, \quad (3)$$

where $\nabla^2$ denotes the Laplacian operator, and $A(\mathbf{r})$ is the distribution of absorbed optical energy density. The constants $\beta$, $c$ and $C_p$ denote the thermal coefficient of volume expansion, speed of sound and the specific heat capacity of the medium at constant pressure, respectively. For this work, the image reconstruction requires the coordinates of the transducer elements, which depend on the position of the four SUTAs when receiving the PA wave. The transducer elements' coordinates of the SUTA can be calculated as:

$$C_x = R \cdot \cos\theta + \frac{a}{2} - a_0$$
$$C_y = R \cdot \sin\theta + \frac{a}{2} - a_0 \quad (4)$$

where $C_x$ and $C_y$ denote the transducer element horizontal and vertical coordinates in the Cartesian coordinate system. $R$ is the SUTA transducer elements distribute radius, $\theta$ is the transducer elements distribute angle, $a$ is the distance of the SUTAs' distribute as shown in Fig. 3(b) and $a_0$ is a constant of SUTA limited view. In this design, the parameter of $R$ is 60 mm and $a_0$ is 10 mm. By applying the formulate (4), the coordinate of the transducer array can be calculated for imaging reconstruction.

### III. SIMULATION AND EXPERIMENT

#### A. Size-adjustable ROI k-wave Simulation

*K-wave* is an acoustic toolbox in *MATLAB*, it is designed for time domain acoustic and ultrasound simulation [22]. Therefore, to demonstrate the feasibility of the proposed PAT system with ROI size-adjustable capability, we use the k-wave toolbox to simulate the different situations of the system. For the simulation setup, the SUTA is simulated with 32 sensors and linear distribution on a circular arc, whose radius is 6 mm. Therefore, the simulation setup can well simulate the actual design of the SUTA. There are four SUTA distribution sensors used for signal capture, which are with 7.5 MHz, 5 MHz, 2.5 MHz and 1 MHz center frequency and 60% bandwidth with different radius. The simulated sensor arrays are towards the ROI center. To simulate the proposed size-adjustable PAT imaging, the SUTA sensors formed different sizes of ROI. Three kinds of distribution are simulated with small, middle and large ROI, and the parameter $a$ denoted in Fig. 3(b) is set as *0 mm, 2 mm* and *4 mm*, respectively.

A vascular model image is used as a numerical phantom to generate PA signals by the *k-wave* toolbox simulation. To better simulate real PA signals, we assign 40 dB signal to noise ratio, and with a 100 MHz sampling rate. Fig. 5 shows the simulation results. A vessel shape numerical phantom is used for initial pressure distribution simulation as shown in Fig. 5(a). Fig. 5(b), (d) and (f) are the proposed size-adjustable PAT system's simulation results with different sizes. The black dots are the simulated sensor distribution. Fig. 5(b) shows the imaging result, whose parameter *a=0*. It seems comparable to the quality of the imaging result with conventional standard circle distribution, as shown in Fig. 5(c). For the middle size SUTA simulation result, as shown in Fig. 5(d), the parameter *a=2 mm*. The outline of the phantom in this result is distinguishable. Compared with the result in Fig. 5(e), Fig. 5(d) show some artifacts due to the signal reception gap between SUTAs. For the large size of ROI simulation, the parameter *a* is set as 4 mm. As shown in Fig. 5(f), the outline of the phantom is also clear and distinguishable. On the other hand, the gap between SUTAs increase when the imaging size increases, which induces more severe artifacts. The simulated sensor of Fig. 5(c), Fig. 5(e) and

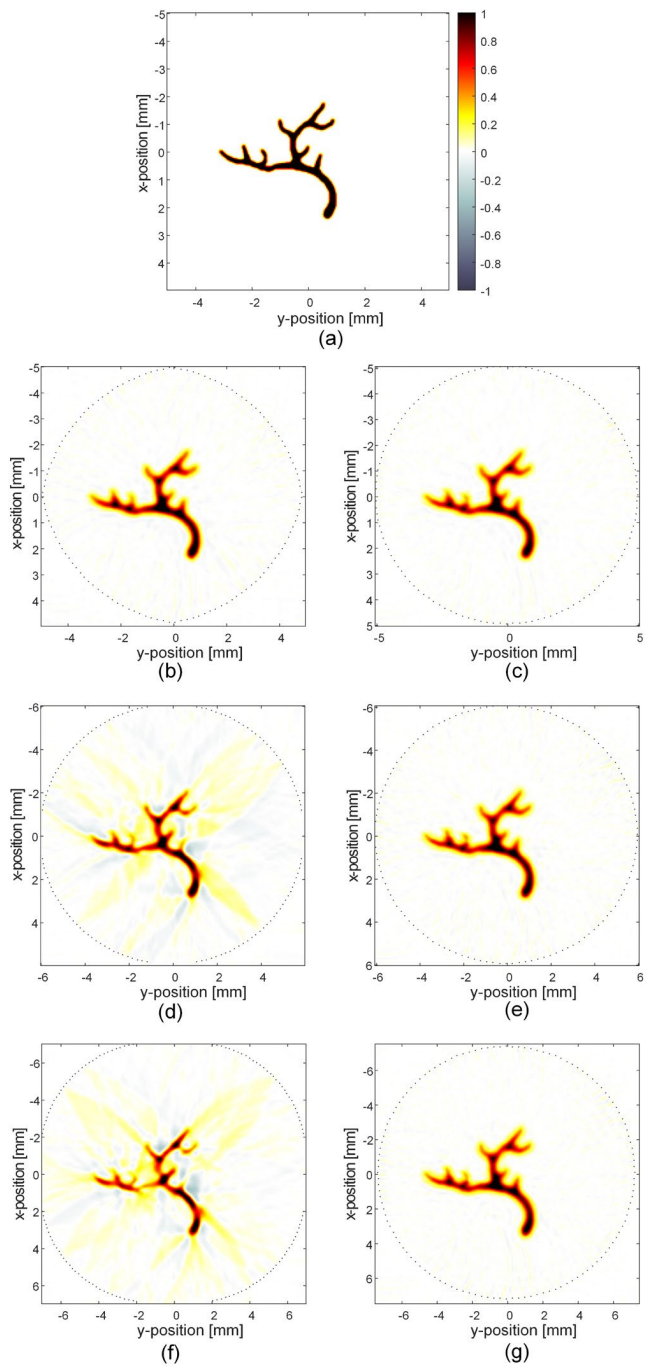

Fig. 5. The simulation results of the size-adjustable PAT system. (a) is the simulated initial pressure distribution of the imaging phantom; (b), (d) and (f) are the simulation results and SUTAs distribution （parameter *a=0 mm*, *a=1 mm* and *a=2* mm） with different size; (c), (e) and (g) are the simulation results for the conventional imaging system setup with circular distributions (radius are 2.5 mm, 3 mm and 3.5 mm).

Fig. 5(g) are with 128 elements and distributed with radius 2.5 mm, 3 mm and 3.5 mm, respectively.

For most instances, the ultrasonic transducer array of the PAT system is nonadjustable with a fixed radius [23]. Therefore, to enhance the compatibility of the PAT system for large target, the conventional full-ring transducer array is designed with a large radius setup. However, due to the acoustic wave attenuation and diffraction, for the small target, the large ring array cannot receive the PA wave very well, compared with the transducer array that is closer to the target [24]. Therefore, the simulations with different size's targets are demonstrated to verify the flexibility and superiority of the SUTAs based ROI size-adjustable PAT system. The simulation phantom is shown in Fig. 5(a) and with different scale changes for small, middle and large phantom simulations. Fig. 6 shows the simulation results of different size imaging targets with conventional full-ring transducer array and SUTAs. Fig. 6 (a), (c) and (e) are the simulation results with conventional full-ring array setup for small, middle and large imaging targets. For the small target, the large ring array received signals cannot reconstruct a clear image as shown in Fig. 6(a) and Fig. 6(b). The edge of the phantom is blur, and some high-frequency information is lost. However, the SUTA-based simulation results shown in Fig. 6(b) and Fig. 6(d) are much clearer and with a sharp outline. For the large imaging target, with the SUTAs outward distribution, the transducer limited view issue is becoming severe. Therefore, the four SUTAs rotate 90 degrees to get a full angle PA signals receiving. Fig. 6(f) shows the simulation result with the four SUTAs rotating 90 degrees that forms a 256 elements UT array, which is better than Fig. 6(e) of conventional full-ring array simulation result.

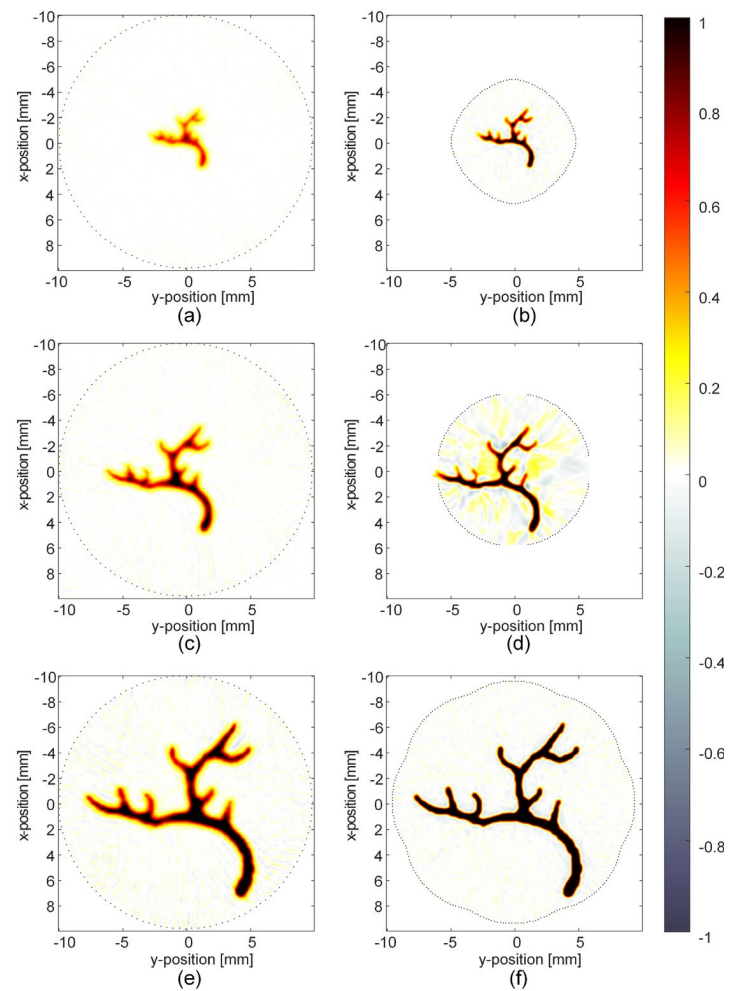

Fig. 6. The simulation results of different size imaging targets. (a), (c) and (e) are the simulation results for small, middle and large imaging targets with conventional full ring transducer array setup; (b), (d) and (f) are the simulation results for small, middle and large imaging target with SUTAs setup.

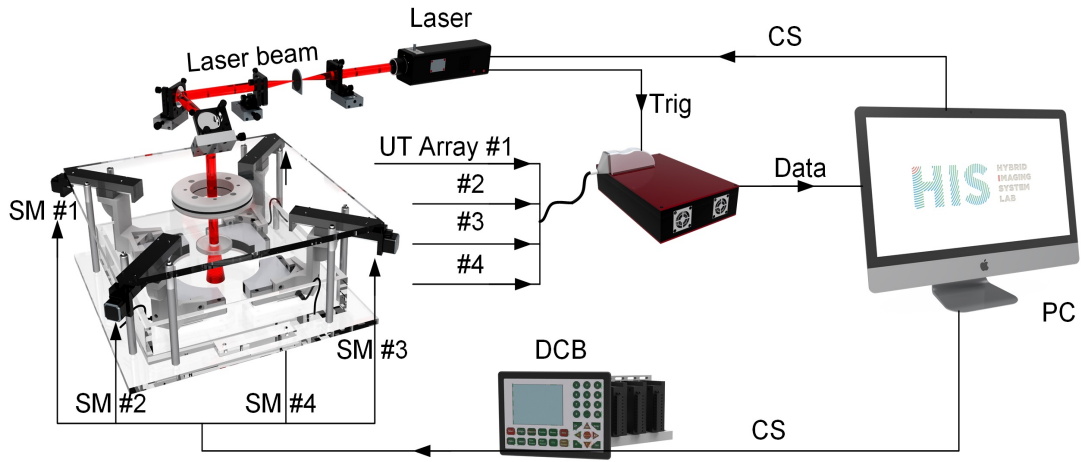

Fig. 7 The proposed size-adjustable PAT system setup. UT: ultrasound transducer; Trig: trigger; CS: control signal; PC: personal computer; DCB: digital control board; SM: stepper motor; #: device number.

The simulation results very well demonstrated the feasibility of the proposed size-adjustable PAT system. The simulated SUTA-based PAT system imaging results are comparable with the results of the conventional PAT system setup, whose sensor array is with the circular distribution. It is worth noting that the proposed SUTA-based size-adjustable PAT system is flexible to address different-size targets by changing the SUTAs distribution, which is not possible for the conventional PAT system setup. Furthermore, there are some artifacts in the imaging results that can be easily reduced by data interpolation and image processing.

### B. The PAT System Setup with SUTAs

The phantom study is executed to demonstrate the feasibility of the proposed size-adjustable PAT system. The PAT system setup with SUTAs is shown in Fig. 7. The laser source (PHOCUS MOBILE, OPOTek Inc. USA) with 10 Hz repetition rate, 750 nm wavelength and 70 mJ energy, is used to excite the phantom to generate PA signal. The four SUTAs form a 128 elements ultrasound receiving array to capture the PA signal. A 128-channel data acquisition (DAQ) module with 40 MHz sampling rate is applied for PA signal sampling (LEGION 128 ADC, PhotoSound Technologies, Inc. USA). There are four stepper motors fixed in a custom-shaped acrylic platform that ensures the four SUTAs towards the imaging center when the motors are moving. A digital control board includes four stepper motor drivers, and a controlling module that controls the movement of the motor. By controlling the movement of the four stepper motors, the SUTAs can dynamically adjust the size of imaging ROI. The imaging target is surrounded by the four SUTAs ensuring that the PA signals can be well detected. The laser homogeneously excited on the imaging phantom after some optical components' adjustment. Meanwhile, the laser clock module outputs the synchronous trigger to the DAQ. After that, the computer reconstructs the PA image by applying universal back-projection algorithms [25, 26].

### C. Size-adjustable PAT Experiments

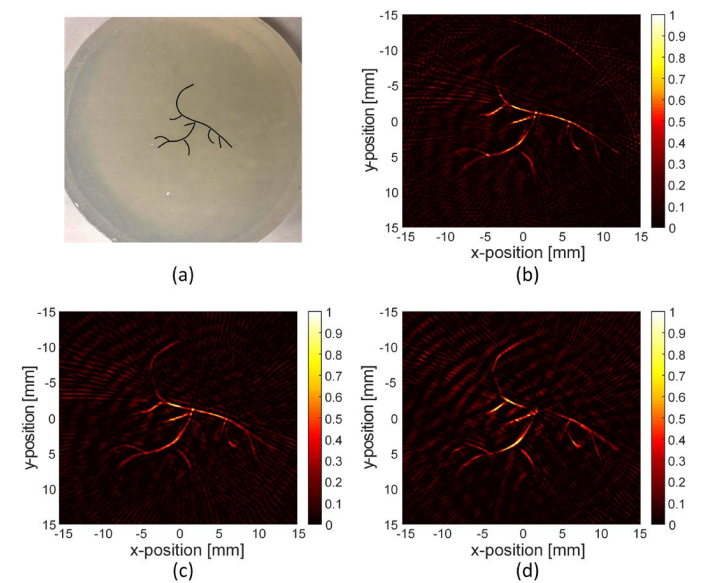

Fig. 8 The phantom imaging results. (a) the photograph of the phantom; (b), (c) and (d) are the imaging results corresponding to *a=0, a=20 and a=30* with different imaging sizes.

To verify the feasibility that the proposed PAT system can adapt to different sizes of imaging targets, i.e., the size of ROI can be adjusted according to the distribution of SUTA, the phantom imaging results with different sizes of ROI are compared. The imaging sample is a blood vessel-like phantom made of black iron wire with 0.3 mm diameter embedded in an agar block as shown in Fig. 8(a). Fig. 8(b), Fig. 8(c) and Fig. 8(d) are the imaging results of the phantom corresponding to

*a=0, a=20* and *a=30* with different imaging ROI. The imaging results show that the PAT system, which is based on the SUTA distribution adjustment to achieve different ROI size, can image the phantom with a clear profile.

The imaging results verified the feasibility of the proposed size-adjustable PAT system to adapt for different sizes' targets by changing the SUTAs distribution. Despite the imaging size increasing, there are more artifacts in the imaging result, and the imaging result can be greatly improved by signals interpolation and image processing[27, 28]. Therefore, the PAT system based on SUTAs is more flexible for different sizes of imaging targets than the conventional PAT system with a fixed circular distribute UT array.

*D. Ex-vivo imaging*

An ex-vivo pork breast with ICG injection is imaged by the proposed PAT system, which verified its feasibility for biomedical applications. The pork breast *ex vivo* phantom composed by skin, fat and capillaries etc., can very well simulate the biomedical application scenarios. Fig. 9(a) and Fig. 9(b) are the photographs of the *ex-vivo* pork breast meat. The contrast agent ICG (Dandong Yichuang Pharmaceutical Co. Ltd.) is injected at a depth of 2 to 3 cm under the skin of the phantom. Fig. 9(b) shows the lateral view of the phantom after injection. The ICG is an FDA-approved contrast agent for PA signal enhancement in deep tissue. For the imaging setup, the wavelength of the laser is 780 nm that matches the ICG's maximum absorption wavelength. The fluence of the laser is less than 20 mJ/cm$^2$ at the skin surface.

Fig. 9(c) shows the ICG injected pork breast meat imaging result, there is a bright target with high contrast to the background that is the injected ICG. This result shows the feasibility of the proposed system for tissue imaging.

IV. CONCLUSION

In this paper, the size-adjustable PAT system is proposed that can adapt for different sizes' imaging targets. This system greatly improves the flexibility of the PAT system for biomedical applications. Specifically, the proposed system can apply for human breast cancer imaging with different sizes to achieve optimum imaging performance. The SUTAs with different center frequencies can detect the PA signals with higher fidelity. The size-adjustable PAT system is based on the SUTAs distribution that can be motorized to adjust the imaging ROI. The imaging size of the system is based on the control of four stepper motors' movements that adjust the SUTAs distribution. The four SUTAs can adjust the imaging ROI's radius ranging from 50 mm to 90 mm. A calibration processing of the PA signals should be carried out before applying the back-projection algorithms to reconstruct the image.

Simulation and experimental results of a vascular model have well demonstrated the feasibility of the proposed PAT system with the SUTAs. Moreover, further advanced data processing algorithms can be used to improve imaging quality. The proposed system shows high potential to promote the PAT system for wider clinical applications with enhanced flexibility.

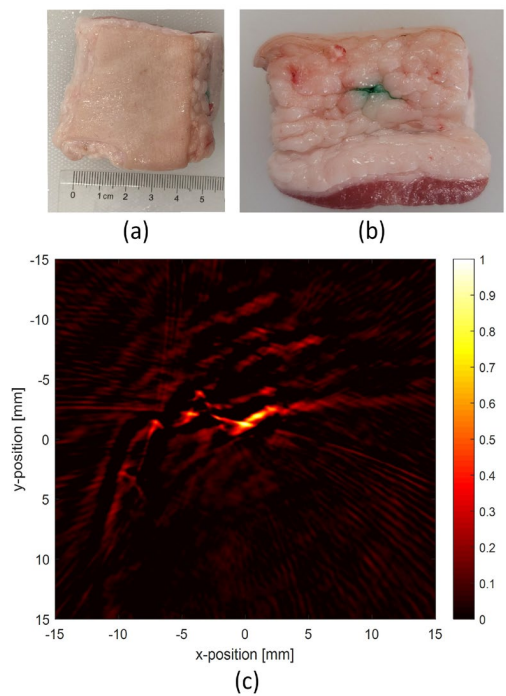

Fig. 9 The *ex-vivo* imaging results. (a) the photograph of the pork breast; (b) the lateral view of the pork breast with ICG injections; (c) the imaging result.